\documentstyle[12pt]{article}
\title{The past and present infrared spectrum of
BD+30$^\circ$3639.}
\author{Albert A. Zijlstra and Ralf Siebenmorgen\\
\\
\\
European Southern Observatory\\
Karl-Schwarzschildstr. 2\\
D-85748 Garching bei M\"unchen\\
Federal Republic of Germany\\
\\
\\
\\
Paper presented at the workshop:\\
 Planetary Nebula Nuclei:
Models and Observations,\\
Bachotek 31 Augustus -- 2 September 1993
\\
\\
to be published in Acta Astronomica\\
}
\newpage
\begin{document}
\maketitle
\newpage


\def\wisk#1{\ifmmode{#1}\else{$#1$}\fi}
%
%
\def\k{\,{\rm K}}
\def\khz{\,{\rm kHz}}
\def\mhz{\,{\rm MHz}}
\def\ghz{\,{\rm GHz}}
\def\kmsmpc{{\rm\,km\,s^{-1}\,Mpc^{-1}}}                       
\def\jy{\,{\rm Jy}}                                            
\def\kpc{\,{\rm kpc}}                                          
\def\pc{\,{\rm pc}}
\def\yr{\,{\rm yr}}                                            
\def\mjy{\,{\rm mJy}}
\def\magg{\,{\rm mag}}                                         
\def\peryr{\,{\rm yr^{-1}}}
\def\cm{\,{\rm cm}}
\def\cmd{\,{\rm cm^{-3}}}
\def\angstrom{\,{\rm\AA}}
\def\msolar{\,{\rm M_\odot}}
\def\lsolar{\,{\rm L_\odot}}
\def\mum{\mu{\rm m}}
%
%
\def\hbeta{\wisk{ H\beta}}
\def\lhbeta{\wisk{ L_\hbeta}}
\def\fhbeta{\wisk{ F_\hbeta}}
v\def\lyalpha{\wisk{ Ly\alpha}}
\def\halpha{\wisk{ H\alpha}}
\def\ebv{\wisk{ E_{B-V}}}
\def\teff{\wisk{ T_{eff}}}
\def\lstar{\wisk{ L_\star}}
\def\tstar{\wisk{ T_\star}}
\def\lsol{\wisk{\,\rm L_\odot}}
\def\msol{\wisk{\,\rm M_\odot}}
\def\rsol{\wisk{\,\rm R_\odot}}
\def\tsol{\wisk{\,\rm T_\odot}}
\def\kms{\wisk{\,\rm km\,s^{-1}\,}}                    
\def\wcm{\wisk{\,\rm W\ cm^{-2}\,}}                    
\def\wcmmu{\wisk{\,\rm W\,cm^{-2}\,\mu m^{-1}\,}}      
\def\cmcub{\wisk{\,\rm cm^{-3}\,}}                     
\def\tenpow#1{\wisk{\,\rm 10^{#1}\,}}                  
\def\timtenpow#1{\wisk{\,\rm\times10^{#1}\,}}          
\def\jyb{\wisk{\,\rm Jy\ beam^{-1}\,}}                 
\def\mjyb{\wisk{\,\rm mJy\ beam^{-1}\,}}               
%
%
\def\gt   {$\!$\hbox{\tt >}$\!$}
\def\lt   {$\!$\hbox{\tt <}$\!$}
\def\oversim#1#2{\lower0.5pt\vbox{\baselineskip0pt \lineskip-0.5pt
     \ialign{$\mathsurround0pt #1\hfil##\hfil$\crcr#2\crcr\sim\crcr}}}
\def\gsim{\mathrel{\mathpalette\oversim>}}    
\def\lsim{\mathrel{\mathpalette\oversim<}}    
\def\deg{\wisk{^{\rm o}}}                                
\def\degpt#1{#1\wisk{^{\rm o}}}                         
\def\arcmin#1{#1\wisk{^{\prime}}}                        
\def\arcsec#1{#1\wisk{^{\prime\prime}}}                  
\def\decdeg#1.#2{\wisk{#1^{\,\rm o}\bck.\,#2}}

\def\sdecdeg#1.#2{\wisk{#1^{\phantom{\,\rm o}}\bck.\,#2}}
%

%
%

\begin {center}
{\Large \bf The past and present infrared spectrum of
BD+30$^{\circ}$3639.}
\end {center}
\author{Albert A. Zijlstra and Ralf Siebenmorgen
\\
\\
European Southern Observatory\\
Karl-Schwarzschildstr. 2\\
D-85748 Garching bei M\"unchen\\
Federal Republic of Germany}
\vskip2.0truecm
\begin{abstract}

We present a radiative-transfer calculation which reproduces the
infrared spectrum of the planetary nebula BD+30$^{\circ}$3639,
fitting both the spectral energy distribution and the spatial extent
at infrared wavelengths. We obtain an acceptable fit to most of the spectrum,
including the infrared bands.  The fit requires a distance of $\ge 2
\kpc$, which implies that BD+30$\deg$3639 has evolved from a massive
progenitor of several solar masses.  Two surprising results are (1) a
very low dust-to-gas ratio, (2) an absence of the smallest PAH
molecules.  Extrapolating back in time, we calculate the previous
infrared evolution of BD+30\deg{}3639.
\\
\end{abstract}

\section{Introduction}

BD+30$\deg$3639 is one of the best studied planetary nebulae (PN) in
the infrared. We have used a recently developed dust model, which
incorporates all the important dust components in the interstellar
medium, to fit its infrared spectrum. Such calculations allow us to
study the dust composition in PN, which significantly differs from
that in the interstellar medium.  Also, we can extrapolate the model
back in time to study how the infrared spectrum may have evolved after
the end of the mass loss.  We have to solve two basic problems: i) The
objects are highly embedded in an optically thick dusty envelope and
the equation of radiative transfer has to be properly handled. ii) A
more or less realistic description of the nature of interstellar dust
particles is required as well.

In this paper we will first present the dust model.
We than fit the spectral energy distribution of the
carbon-rich PN BD+30$\deg$3639.  The calculated spectra at
several stages of the preceding post-AGB evolution are presented in
Section 4.

\section{The dust model}

The dust model used in the present calculations is that of
Siebenmorgen \& Kr\"ugel (1992, henceforth SK).
The dust model is governed by a grain size distribution
($n(a)\propto a^{-q}$) with radii between approximately 2500\AA \/
down to molecular sizes of about 5\AA \/ ($\sim$ 25 atoms) in which we
distinguish three different populations of dust particles:
\\
i) Large particles with sizes $a \geq 100$\AA, $q = 3.5$, causing the
far-infrared/submillimeter emission.  These particles have sufficient
large heat capacities that we can neglect the quantum-statistical
behaviour of photon--grain interactions.  We have used the optical
constants for amorphous carbon.
\\
ii) Small graphite grains with sizes $10\AA \leq a <
100\AA$, $q = 4$, emitting predominately in the mid-infrared.
These particles show temperature fluctuations after individual energy
absorption events (e.g. with photons, electrons); to calculate their
emission spectrum one has to consider multi-photon events.
\\
iii) Small PAHs ($\sim 25$ atoms) and larger PAH clusters ($\sim 250$
atoms), which are the probable carriers of the family of emission
features between $3 - 14 \mum$ (e.g.  Allamandola et al. 1989a).
The ratios of the various bands is determined by the ydrogen-to-carbon
atom ratio and by the size distribution of these molecules.

The solution of the radiative transfer problem including scattering
for spherically symmetric dusty objects containing quantum-heated
particles is described in Siebenmorgen et al.  (1992).
The heating source is described as a
blackbody with an adopted stellar temperature $T_{*}$ and bolometric
luminosity. The density distribution is defined here by means of power
laws $n(r) \propto r^{\beta}$.

\section{The infrared spectrum of BD~+30\deg{}3639}

Our fit is shown in Figure 1.  The data points are a literature
compilation; the lower panel shows new data of Schutte and Tielens
(1993).  We have used a distance of $2\kpc$, which is the smallest
distance for which we were able to obtain a satisfactory fit of the
infrared spectrum.  Hajian and Terzian (1993), from a measurement of
the radio expansion, find a distance of $2.68\pm0.81\kpc$.  The
density structure is modelled as an inner region of uniform density
around a central cavity, surrounded by a larger region where the
density falls as $r^{-2}$.  We assume that the dust composition does
not depend on position in the nebula.

The fit is accurate over most of the spectrum to within $\leq 20\%$,
although one should take in consideration the simple density model
used and the uncertainties in the optical constants of the dust
properties.  The observations become noisier near $5\mum$, which
presents a problem in where to define the continuum level.  The PAH
features at 7.7 and $8.6\mum$ are not well fitted.  For the latter,
the SK model consistently predict values too low by about a factor
$\leq 3$ and it is likely that the cross section used for this feature
is too low.  The feature at $7.0\mu$m, which is not fitted, is due to
an [ArII] line.  The last two data points have a large contribution
from free--free emission.  The measured stellar flux at K (0.09Jy)
agrees very well with the model prediction (0.1Jy).

Radial profiles at several wavelengths have been published by Hora et
al. (1993).  Our calculations reproduce these well.  However, we find
that the predicted radial flux distribution is almost constant between
3 and 15$\mum$, whereas Hora et al. find that the PAH features are slightly
more extended than the 10$\mum$ continuum. This could be due to the
destruction of PAHs in the ionized region by Ly$\alpha$ emission.

We find a dust-to-gas ratio of $3.5\times10^{-3}$, confirming earlier
findings, e.g. Hoare et al. (1993). This is a surprisingly low ratio,
well below the generic value for the interstellar medium. PN such as
BD+30$^{\circ}$3639 do not appear to enrich the ISM with dust.

The PAH emission is completely dominated by the largest
PAH component, the cluster molecules. The contribution from the small
PAHs is almost negligible.  The band ratios of BD+30$^{\circ}$3639
yield a low H/C ratio of the PAH clusters.  The 11.3$\mu$m feature can
also be identified with SiC (e.g.  Hoare et al.  1992).  We failed to
reproduce the spectrum assuming that the 11.3$\mu$m feature is
completely due to SiC, and conclude that the SiC contribution is
likely $<50\%$.

We predict a significant extinction towards the central star, caused
by the dust inside the nebula: $A_{\rm V}=0.43$.  The assumption that
the extinction towards the star and the nebula are equal does not
appear to be validated for BD+30$^{\circ}$3639.

The model predicts a stellar luminosity of $1.3\times 10^4\lsolar$.
We have used the lowest acceptable distance, and therefore this should
be taken as a lower limit.  The luminosity--core mass relation
(e.g. Boothroyd and Sackman 1988) implies a corresponding core mass of
0.7--0.75$\msolar$.  Of the well-known galactic PN, only NGC~7027 and
NGC~6369 are known to have such high luminosities (Gathier and
Pottasch 1989), where it should be noted that NGC~7027 appears to have
entered the cooling track in the HR diagram, and therefore will have
had a higher luminosity in the past.  The total nebular mass in our
fit is about $2.5\msolar$. This value is poorly determined, however it
clearly also implies a massive progenitor of several solar masses,
which has shed a large amount of mass. From the inner radius of the
nebula, and using the known expansion velocity, we derive an age of
the nebula of around $600\yr$, measured from the end of the mass-loss
phase.  This short transition time also implies a high-mass star.

\section {The infrared evolution of BD+30$^{\circ}$3639}

The density distribution used in the final fit allows one to calculate
the preceding infrared evolution of BD~+30$^{\circ}$3639.  In
calculating the progenitor evolution we have made a number of
simplifying assumptions. First, the entire nebula was represented only
by the $1/r^{-2}$ dust density component. We extended the wind-like
component of the fit inward, keeping the total nebular mass constant.
Second, we calculated the age of BD+30$^{\circ}$3639 from the adjusted
inner radius and the observed expansion velocity of $22\kms$.  Third,
we assumed that the mass loss terminated at a stellar temperature of
$5000\k$, and that the stellar temperature has increased linearly
during the post-AGB evolution.  The first step of the evolutionary
sequence is calculated when the inner radius reaches $10^{15}\cm$,
approximately ten years after the mass loss ends. In this way we avoid
the problem at what distance from the star the dust particles form.

Figure 2 shows six evolutionary steps.  The first step exhibits a
steep cut-off around $2.5\mu$m, caused by the high optical depth in
the circumstellar envelope. At this phase the visual extinction will
drop quickly, and the star will become visible at near-infrared
wavelengths within $25\yr$. Optical visibility follows after $\sim
100\yr$. The submillimetre continuum slowly increases with time,
caused by the fact that the dust temperature in the outer regions of
the cloud increases as the optical depth towards the star goes down.
The PAH features develop quickly when the star is still relatively
cool, and reach full strength around $15000\k$.

Figure 3 presents the evolution of the IRAS colours in the
post-mass-loss phase. The colours evolve significantly during the
first 500yr, after which the evolution slows down. At that stage the
colours are already close to the colours exhibited by planetary
nebulae.  Thus, this implies that slowly-evolving transition objects
will have colours close to those of PN.  Surveys which concentrated on
objects with colours more similar to AGB stars will have been biased
in favour of young post-AGB objects.

\vskip 33pt
\noindent {\bf References}
\begin{list}
{}{\itemsep 0pt \parsep 0pt \leftmargin 3em \itemindent -3em}

\item Acker A., Ochsenbein, F., Stenholm, B., Tylenda, R., Marcout,
J., Schon, C. 1992, Stras\-bourg--ESO catalogue of galactic planetary
nebulae (European Southern Observatory, Garching)

\item Allamandola L.J., Tielens, A.G.G.M., Barker, J.R., 1989a, ApJS,
71, 733

\item Boothroyd A.I., Sackman, I.J. 1988, ApJ 328, 641

\item Gathier, R., Pottasch, S.R., 1989, A\&A 209, 369

\item Hajian, A.R., Terzian, Y., Bignell, C. 1993, NRAO preprint

\item Hoare M.G., Roche, P.F., Clegg, R.E.S. 1992, MNRAS 258, 257

\item Hora J.L., Deutsch, L.K., Hoffmann, W.F., Fazio, G.G.,
Shivanandan, K. 1993, ApJ, 413, 304

\item de Muizon M.J., d'Hendecourt, L.B., Geballe, T.R., 1990, A\&A
227, 526

\item Schutte W.A., Tielens, A.G.G.M., 1993, ApJ, in press

\item Siebenmorgen R., Kr\"ugel, E.,1992a,A\&A, 259, 614, (SK)

\item Siebenmorgen R., Kr\"ugel, E., Mathis, J.S., 1992, A\&A 266,
501

\end{list}

\pagebreak

\noindent {\bf  Figure Captions}
\begin{list}
{}{\itemsep 0pt \parsep 0pt \leftmargin 3em \itemindent -3em}

\item Figure 1: a) The spectral energy distribution of
BD+30$^{\circ}$3639.
Observations are shown by squares, the model
with amorphous carbon grains for the large-grain component is depicted
by a solid line.
 b) High-resolution spectrum  between 3
and 14$\mu$m.  Observations shown are from Schutte and
Tielens (1993).

\item Figure 2: The infrared time evolution between 0.1 and 1000$\mu$m of
BD~+30$^{\circ}$3639. The first evolutionary sequence is calculated 13
years after the mass loss ends.

\item Figure 3: The calculated evolution of the IRAS colours.
The open circles are IRAS data of planetary nebulae.
\end{list}

\end{document}